\documentclass[letterpaper,referee]{raa}            

\usepackage{graphicx,times}             
\usepackage{natbib}
\usepackage{amssymb,amsmath}
\bibpunct{(}{)}{;}{a}{}{,}

\usepackage[letterpaper]{geometry}
\usepackage[pagebackref=true]{hyperref}
\hypersetup{colorlinks = true, linkcolor = green, anchorcolor = red, citecolor = blue, filecolor = red, urlcolor = red}

\begin{document}

   \title{Development of Technique to Detect and Classify Small-Scale 
Magnetic Flux Cancellation and Rapid Blueshifted Excursions
}

   \volnopage{Vol.0 (200x) No.0, 000--000}      
   \setcounter{page}{1}          

   \author{Xin Chen
      \inst{1}
   \and Na Deng
      \inst{1}
   \and Derek A. Lamb
      \inst{2}
   \and Ju Jing
      \inst{1}
   \and Chang Liu
      \inst{1}
   \and Rui Liu
      \inst{3}
   \and Sung-Hong Park
      \inst{4,5}
   \and Haimin Wang
      \inst{1}
   }

   \institute{Space Weather Research Laboratory, New Jersey Institute of Technology, University Heights, Newark, NJ 07102-1982, USA; {\it xc55@njit.edu}\\
        \and
             Southwest Research Institute, 1050 Walnut Street, Suite 300, Boulder, CO 80302, USA\\
        \and
             CAS Key Laboratory of Geospace Environment, Department of Geophysics and Planetary Sciences, University of Science and Technology of China, Hefei, Anhui 230026, China\\
		\and
			 Institute for Astronomy, Astrophysics, Space Applications and Remote Sensing, National Observatory of Athens, Penteli 15236, Greece\\
		\and
			 Korea Astronomy and Space Science Institute, Daejeon 305-348, Republic of Korea\\
   }

   \date{Received~~2009 month day; accepted~~2009~~month day}

\abstract{We present a set of tools for detecting small-scale solar magnetic cancellations and the disk counterpart of type II spicules (the so-called Rapid Blueshifted Excursions (RBEs)), using line-of-sight photospheric magnetograms and chromospheric spectroscopic observations, respectively. For tracking magnetic cancellation, we improve the Southwest Automatic Magnetic Identification Suite (SWAMIS) so that it is able to detect certain obscure cancellations that can be easily missed. For detecting RBEs, we use a normalized reference profile to reduce false-positive detections caused by the non-uniform background and seeing condition. Similar to the magnetic feature tracking in SWAMIS, we apply a dual-threshold method to enhance the accuracy of RBE detection. These tools are employed to analyze our coordinated observations using the Interferometric BIdimensional Spectrometer at Dunn Solar Telescope (DST) of the National Solar Observatory (NSO) and \textit{Hinode}. We present the statistical properties of magnetic cancellations and RBEs, and explore their correlation using this data set.
}

   \authorrunning{X. Chen et al.}            
   \titlerunning{Detect and Classify Magnetic Flux Cancellation and RBEs }  

   \maketitle

%
%
\section{Introduction}           
\label{sect:intro}

Observations of small-scale dynamics in the solar chromosphere have benefited from significant improvements in instrumentation in the past decade. In the earlier on-disk H$\alpha$ observations, one kind of small-scale feature was discovered to show blue-shifted (upflow) component only with no corresponding red-shifted (downflow) component. They were named H$\alpha-1.0$~\AA\ jets \citep{Wang98} or chromospheric upflow events \citep{Chae98,CYLee00}. It was found that these jets tend to occur at supergranular boundaries, and sometimes recur on the same sites. They mostly appear to be round in shape rather than elongated. More recently, when observed at the solar limb, the highly dynamical type II spicules were distinguished by their outward-only ejection, high speed (15--40 km/s), and short lifetime ($<$150 s) comparing to the ``classical'' type I spicules \citep{DP07}. Later, these type II spicules were linked to their on-disk counterparts, the Rapid Blueshifted Excursions (RBEs; \citealp{Lang08,RvdV09}), which are essentially similar to the upflow events mentioned above.

Using observations with improved resolution, more detailed properties of RBEs were revealed. Statistically, their occurrence rate is compatible with that of the type II spicules \citep{RvdV09,Sekse12}. In addition, RBEs are mostly elongated and their upflows are accelerated from the footpoint to the top end. Although they have been known separately in H$\alpha$ and Ca II 8542~\AA\ observations for several years, new studies show that their positions and accelerations exhibit consistency from the lower layer (Ca II) to the higher layer (H$\alpha$) of the chromosphere \citep[e.g.,][]{Sekse12}. Sometimes, this connection can also be extended to corona to appear as bright points according to observations made by the Atmospheric Imaging Assembly (AIA) on board the Solar Dynamics Observatory (SDO) \citep{DP11}. Besides the upflow motion in the line-of-sight (LOS), RBEs also show transversal and torsional motions \citep{Sekse13+}, which are of comparable magnitude to the type II spicules \citep{DP12}. Furthermore, features in the red wing but with similar characteristics to RBEs were discovered and named Rapid Redshifted Excursions (RREs; \citealp{Sekse13+}). It was found that RREs appear less frequent than RBEs, especially near the disk center. A significant fraction of RREs occur together with RBEs, which is interpreted by \citet{Sekse13+} as being due to the upflow, transverse, and torsional motions in combination with certain viewing angles.

To study the properties of the numerous small-scale dynamic features like RBEs, an appropriate detecting and tracking method would be important. Since RBEs are tiny (few arcseconds) and ephemeral and display diverse spectral profiles, detecting them requires observations with high spatial, temporal, and spectral resolution. There is an automatic algorithm developed for the RBE studies \citep{RvdV09,Sekse12}, which is designed for observations taken by the Crisp Imaging Spectropolarimeter at the Swedish Solar Telescope on La Palma. It combines different methods to detect RBEs, such as using Doppler images derived from the difference between the blue- and red-wing images, and using the extreme far blue-wing image as a reference to eliminate background. This algorithm requires multi-frames restoration of images as well as excellent seeing; otherwise, the mismatch of features in different wavelengths would induce many false-positive detections during subtraction. In this study, we develop a new automatic RBEs tracking algorithm, which has a better tolerance of minor mismatches induced by image distortions or occasional seeing variations. Moreover, when the nearly simultaneous photospheric magnetograms are available, our tool as described below provides properties of RBEs and the comparison results with their associated photospheric magnetic features.

As almost all solar activities are related to the magnetic field, the magnetic configuration and dynamics associated with RBEs have been investigated in many studies. It is known that even in the quiet-Sun region, the magnetic field is not tranquil but has dynamic network and intra-network structures. As mentioned previously, RBEs tend to occur near the concentration of photospheric magnetic field, and are sometimes associated with converging magnetic dipoles \citep{Wang98}. Considering their highly dynamic characteristics, RBEs may be propelled by small-scale magnetic reconnection. In this case, magnetic cancellations are likely to indicate the source locations of RBEs. On the other hand, using the three-dimensional (3D) MHD simulations, \citet{MS11} showed that small-scale flux emergence would trigger a chromospheric jet similar to type II spicules. The plasma in the chromosphere can be heated and accelerated by a strong, mostly horizontal Lorentz force and is then ejected along the vertical magnetic field. Correspondingly, there is a case study showing that RBEs are related to the newly appeared magnetic flux concentrations \citep{Yur13}. However, the flux emergence and cancellation are likely to associate with each other. When new flux concentrations emerge from the bottom of the photosphere, they may cancel with opposite polarity fields in the vicinity. Thus far, although the importance of the RBE-associated magnetic field evolution has been generally recognized, the driving mechanism of RBEs is still under active investigation.

In order to study the magnetic configuration and evolution related to RBEs, we take advantage of a well-developed and widely used solar magnetic field tracking method, the Southwest Automatic Magnetic Identification Suite (SWAMIS; \citealp{DF07}), which can track weak-field features close to the noise level. Applying SWAMIS to LOS magnetograms taken by the Michelson Doppler Imager on board the \textit{Solar and Heliospheric Observatory} and the Solar Optical Telescope (SOT; \citealp{Tsuneta08}) of \textit{Hinode}, \citet{Lamb08,Lamb10} demonstrated one kind of observational effect: sometimes the apparent unipolar magnetic flux emergence is actually the coalescence of tiny, previously existing fluxes, which can be observed in higher resolution. Recently, they also found a similar process when unipolar magnetic features disappear \citep{Lamb13}, meaning that the dispersal of flux concentrations might play a more important role than bipolar cancellations in the quiet Sun. However, we found that SWAMIS is likely to miss many cancellations under certain conditions, especially when the sizes of canceling features are significantly different. In order to better detect and characterize the photospheric magnetic cancellation, we thus developed an algorithm that specializes in tracking cancellations using the intermediate results of SWAMIS. The location and timing of the detected flux cancellations can then be compared with those of the detected RBEs.


\section{Observations}
\label{sect:Obs}

We carried out a coordinated observing campaign to obtain quasi-synchronous photospheric magnetograms and chromospheric spectral images during 2011 October 16--23. The LOS magnetograms were obtained by the Narrowband Filter Imager (NFI) of \textit{Hinode}/SOT using the Na D 5896~\AA\ line, with a field-of-view (FOV) of 131$''\times$123$''$, a cadence of 64~s, and a pixel size of 0.16$''$~pixel$^{-1}$.

The spectral data of H$\alpha$ and Ca II lines were acquired from the Interferometric BIdimensional Spectrometer (IBIS; \citealp{Cavallini06}) at the Dunn Solar Telescope (DST) of the National Solar Observatory (NSO; \citealp{Zirker98}), Sacramento Peak. IBIS is equipped with a dual Fabry-P$\acute{e}$rot interferometer system. For H$\alpha$ observation, each scan contains 28 points between $-$1.7~\AA\ to 1.0~\AA\ relative to 6562.8~\AA, and for Ca II, it contains 34 points from 8540.2~\AA\ to 8543.5~\AA. Besides the spectral scans (narrowband channel), we also took synchronous white-light (broadband channel) images for alignment and calibration purposes \citep{Cavallini06,R&C08}. Both cameras have 100$''\times$100$''$ FOV (with an effective 48$''$ radius circular area), and run at speed of $\sim$6 frames second$^{-1}$ with 30~ms exposure time. As a result, the cadence for a full scan is 4--5~s for H$\alpha$ and 5--6~s for Ca II. The image scale is 0.1$''$~pixel$^{-1}$, while the spatial resolution varies with seeing conditions.

During our observation run, we carefully maintained the consistency of pointing of both instruments for a maximal common FOV. Different types of regions (e.g., quiet Sun, active regions, coronal holes) were observed. The data set presented in this paper was targeted at a quiet-Sun region near the disk center, which was observed during 14:30--15:30 UT using H$\alpha$ on 2011 October 21. The Ca II observation of the same day is also elaborated in a separate study \citep{Deng14}. The atmospheric seeing at DST during this time period was moderate to very good. With the help of the Adaptive Optics, the image quality is acceptable as the sub-arcsec resolution is achieved. 

\section{Methods}
\label{sect:Meth}


\subsection{Tracking Photospheric Magnetic Field}
\label{subsect:smt}

\begin{figure}
   \centering
   \includegraphics[width=14.0cm, angle=0]{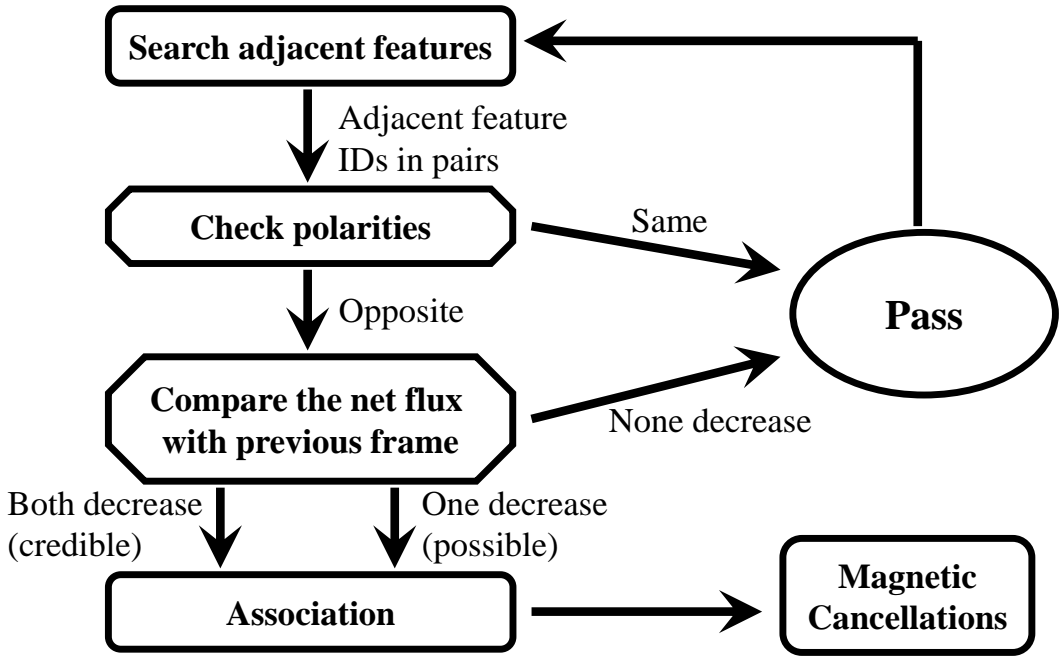}
   \caption{Workflow of tracking magnetic cancellation using a modified code based on SWAMIS.} 
   \label{Fig1}
   \end{figure}
   
For the magnetograms in our coordinated observation, the main goal is to find the properties of magnetic cancellation, which appears as a reduction of net flux of the adjacent opposite-polarity features. Since RBEs are tiny activities, detecting them requires a method that is capable of accurately characterizing small-scale magnetic elements in the weak field regions, such as quiet Sun and coronal holes. As mentioned in Section~\ref{sect:intro}, SWAMIS has demonstrated to be able to automatically track magnetic features close to the noise level \citep{DF07}. 

As a brief introduction, SWAMIS works in five steps on data that has typically been preprocessed to reduce the noise floor and remove perspective effects:
(1) Discrimination: for each frame, determine regions of potential features;
(2) Identification: for each frame, index potential features in marked regions;
(3) Association: connect features across different frames;
(4) Filtering based on size/longevity: remove occasional cluster of noise;
(5) Classification of origin and demise \citep{DF07}.
As a result, SWAMIS provides properties of each magnetic feature in each frame (flux, area, location), and summary information for each feature (birth and death times and the ways of birth and death).

After a careful investigation, we found that SWAMIS often misses cancellations in a particular situation, when the cancellation occurs between two opposite-polarity features with significantly different sizes. In this case, one magnetic feature has considerably more unsigned flux than the other. During such a cancellation event, the larger feature could even show an increase of unsigned flux due to random noise accumulated within the extended area, or due to its simultaneous merger with other like-polarity features. This presents a challenge to SWAMIS, as it tests the flux balance to confirm a cancellation event, under the assumption that the unsigned flux would decrease for both features involved as in a standard flux cancellation. As a consequence, this kind of cancellations would be classified as ``Error'' by SWAMIS rather than ``Cancellation'' because of the apparently unbalanced change in flux. In addition, SWAMIS only checks the status of both features when either of them dies out. But sometimes a cancellation event does not entirely eliminate either features; both of them just become weaker and smaller, and then separate from each other. Since both features still exist, SWAMIS would not check their status and consequently would miss these cancellations. For the difficulties mentioned above, we made our cancellation tracking tool mainly by amending the corresponding portion in the original SWAMIS algorithm in two aspects. First, the size ratio of canceling features is now considered in order to define the confidence of their fluxes (see below). Second, the cancellation is no longer treated as the demise of magnetic elements but is monitored during the whole lifetime of the involved features.

In practice, our procedure of tracking magnetic cancellation is described as follows. We first apply the original SWAMIS code to detect magnetic features following the method in \citet{Lamb10}. In preprocessing, the data are calibrated, carefully aligned, and spatially and temporally smoothed using Gaussian kernels (see Section~\ref{subsect:data}). Based on the noise level, image resolution, and data cadence, we then set appropriate thresholds of feature intensity, size, and lifetime. The result of SWAMIS contains the properties of each detected magnetic feature, such as location, flux, and size. We then extract these properties from SWAMIS results and feed them as input to our cancellation tracking tool, which includes the following main steps (Figure~\ref{Fig1}):

i)\ \ \ \ For each frame, search adjacent features (i.e., separated less than 3 pixels, 0.5$''$) and return their ID in pairs;

ii)\ \ \ Check the polarities of each pair, and remove pairs with the same polarity;

iii)\ \ Check the change of net flux of each feature by comparing its unsigned flux in the current frame with that of the previous frame. Record the credibility of cancellation as ``credible'' (if the unsigned fluxes of both features in a pair decrease), ``possible'' (if the unsigned flux of only one feature in a pair decreases), or ``impossible'' (if no feature in a pair exhibits a decrease of the unsigned flux). The pairs labeled with ``impossible'' are removed;

iv)\ \ \ Associate cancellations across different frames in the magnetogram sequence.

The result of our cancellation tracking tool contains analysis of credibility in addition to the general characteristics like flux variation, location, birthtime and deathtime. Quantitatively, in step iii), the ``credible'' cases have a credibility of 1 and ``impossible'' cases correspond to a zero. Those ``possible'' cases have their credibility ranging from 0.25 to 0.75, based on the ratio of sizes (i.e., area) of features. Specifically, as mentioned above, the total flux of a larger feature has a larger absolute uncertainty. For example, if the feature with increased unsigned flux (UF, same-below) is significantly larger (e.g., twice or larger) in size than the other feature with decreased UF, the credibility is set to be 0.75. Similarly, if the UF of larger feature decreases while that of the smaller one increases, the credibility is set to be 0.25. If both features have a similar size, the credibility is set to be 0.5. For each cancellation, the credibility is evaluated in each frame, and the mean value averaged over all frames in which cancellation sustained is used as the final credibility.

Our specialized cancellation tracking tool shows an improvement of detection and also provides more accurate space-time information. Specifically, first it shows a better capability to find cancellations that involve features of different sizes. Using our tool, the flux uncertainty of a large feature would lead to a lower credibility rather than an error, which is more quantitative. Rather than merely checking the last few frames of canceling features, our algorithm thus exploits more information from the data to provide a trustworthy result. Second, our algorithm directly shows the location and initiation time of cancellations, which can not be obtained by SWAMIS as it focuses on magnetic features themselves. For example, central coordinates of the involved magnetic features may not appropriately indicate the cancellation site, when the features have relatively large sizes or irregular shapes. Similarly, the birth time of the involved magnetic features is generally earlier than the onset of cancellations. By handling cancellations as associated events, our results can provide accurate location and time range information of cancellations. Finally, we find that sometimes both features of opposite polarity do not totally cancel out. In general, during a cancellation event, the size of features as well as their unsigned fluxes would decrease. But it does happen that when features become smaller, their remaining parts move apart instead of toward each other. In fact, these features could merge with other like-polarity features or disperse in situ, which are not identified as cancellations by SWAMIS but are recorded by our algorithm. In summary, our modified tool provides more comprehensive and accurate properties of magnetic cancellations.

\subsection{Tracking Chromospheric RBEs}
\label{subsect:rbet}

\begin{figure}
   \centering
   \includegraphics[width=14.0cm, angle=0]{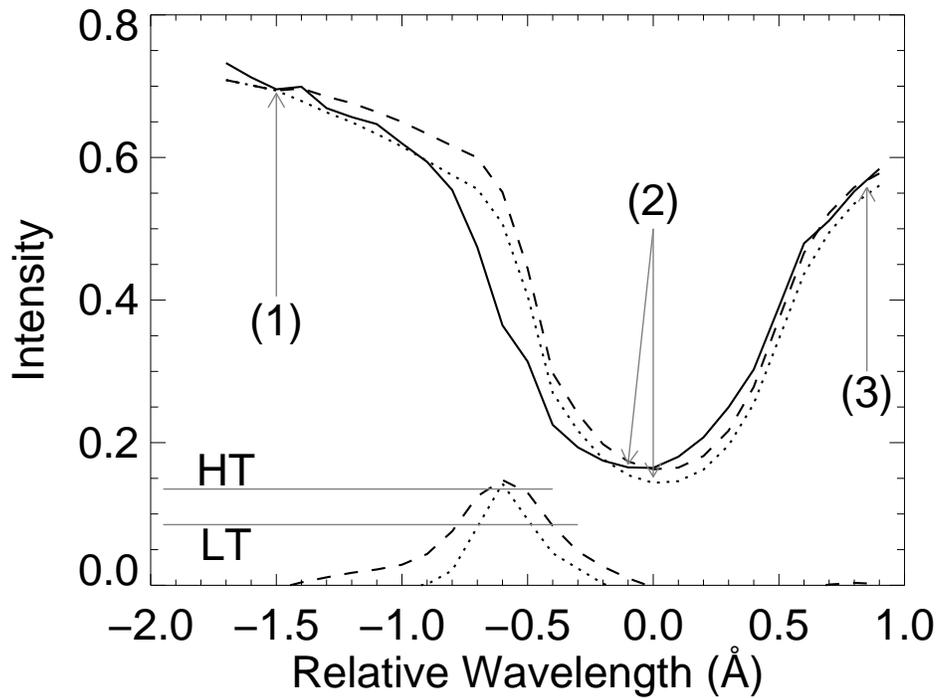}
   \caption{Illustration of normalized reference profile and dual-threshold method. The solid line is the line profile of an arbitrary pixel (local profile). The average profile (dotted line) is multiplied by a scaling factor $f(\lambda)$ to construct the normalized reference profile (dashed line) which has the same intensity of local profile at far blue wing (1), line minimum (2), and far red wing (3). The scaling factor $f(\lambda)$ between points (1)-(2)-(3) is interpolated in order to keep the shape of reference profile. Contrast profiles are plotted at the bottom. The dotted line is derived by the average profile and the dashed line is derived by the normalized reference profile. The horizontal gray lines indicate the high threshold (HT) and low threshold (LT). As an example, this pixel passes the high threshold.} 
   \label{Fig2}
   \end{figure}
   
   \begin{figure}
   \centering
   \includegraphics[width=14.0cm, angle=0]{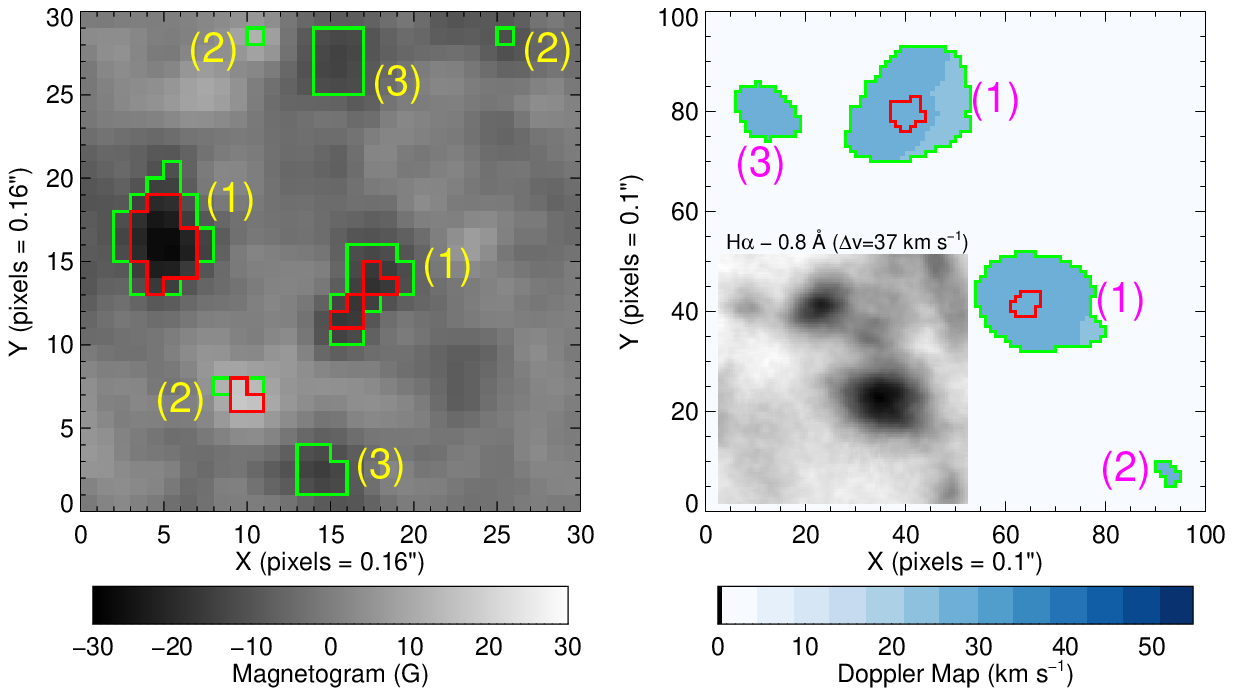}
   \caption{Illustration of dual-threshold method used for tracking magnetic features (left) and RBEs (right). For both panels, the red contours show the high threshold and the green contours correspond to the low threshold. Features labeled (1) have pixels that exceed the high threshold in the current frame; features labeled (2) are small and likely to be filtered out if they are not bigger in some previous or subsequent frame; features labeled (3) are large enough to avoid size-based filtering, do not have any pixels that exceed the high threshold in the current frame, but do have some pixels that exceed the high threshold in some previous or subsequent frame. For the 4D spectral data, we applied the dual-threshold method to the intensity of contrast profiles (see Figure~\ref{Fig2}), which are not shown in the Doppler map (in contrast, thresholds can be directly presented as contours in magnetograms). In the right panel, the insert is an H$\alpha-0.8$~\AA\ image corresponding to this Doppler map (at the same time and location). This gives an intuitive but incomplete view reflecting those thresholds, as only one line position instead of the whole bluewing of H$\alpha$ is presented in the figure.}
   \label{Fig2b}
   \end{figure}

For RBEs, we developed a corresponding automatic tracking method for our four-dimensional ($x$, $y$, $t$, and $\lambda$) data cube observed by IBIS. The task becomes more complicated due to the presence of the spectral dimension. To facilitate the data analysis and achieve our goal, it seems desirable to ``compress'' the wavelength information thus remove the spectral dimension by, for example, making a Doppler map. However, as mentioned above, Doppler maps can not be made directly using the difference between the blue- and red-wing images because of the distortions of IBIS images at different times. Besides, there is a scenario that RBEs and RREs may appear at the same time and location, which would make it not appropriate to use Doppler maps (the blue-shifted component and red-shifted component would cancel each other). We notice that histograms of line-wing images are ``unbalanced'' due to different abundance of RBEs and RREs. In other words, there may exist more regions that show absorptions. Based on the unbalanced histogram as well as the noise level, we try to calculate a threshold for detecting Doppler-shifted regions in the images of each different spectral position. However, this threshold can not distinguish the photospheric background features at far wings, such as granule boundaries. These are the main issues need to be resolved.

In order to properly process the spectral information, our method tries to extract the contrast profile for each pixel with several corrections. Using this contrast profile we can calculate the Doppler velocity and detect RBEs. We describe our method as follows.

First, we normalize reference profiles over the entire FOV to remove the background features. Specifically, rather than using the average profile of the whole FOV as the reference, for each pixel, we scale the average profile to match the local profile but keep the line core unshifted, as illustrated in Figure~\ref{Fig2}. A normalization procedure keeps reference profiles in a similar shape, and allows intensity to vary in different locations to reflect background features (especially at far wings). In other words, normalized reference profiles simulate a profile which has no shifted spectral component at the same local background. In this way, contrast profiles  with most of the background features removed can be constructed (using normalized reference profile minus local profile).

Second, to account for the image distortion, we do a spatial smoothing of images as the multi-frame restoration is not available for our observations. The spatial smoothing process could reduce the noise but it decreases the image resolution as well. For our data set, although the noise level is very low, the distortion is similar to sporadic noise that can be reduced by spatial smoothing. Since the pixel size of IBIS images is much smaller than the resolution, spatial smoothing would in fact not decrease the real resolution of our data set. In addition, as IBIS observation has high spectral resolution, our data set mostly shows a continuous spectral profile. This indicates that any discontinuity, especially the step-like structure in the spectral profile, could be due to a mismatch of pixels in different wavelengths. Therefore we also apply a smoothing procedure to the contrast profile in the spectral dimension. Both the spatial and spectral smoothings can effectively reduce the false-positive detections.
	
Finally, our method utilizes a dual-threshold method to improve accuracy (Figure~\ref{Fig2b}). For solar magnetic field tracking, the dual-threshold method as used by SWAMIS has the advantage in the discrimination of features \citep{DF07}. In its implementation, the high threshold is used to mark the desired pixels, while the low threshold is only applied to pixels that are adjacent to the marked ones. By properly determining both thresholds based on the noise level, this method performs better than detections using a single threshold. However, for the RBE tracking, we can not directly apply this dual-threshold approach to our 4D data cube. For magnetograms, as each pixel has a value that is a superposition of signals (if valid) and noise, the dual-threshold method can effectively extract the LOS magnetic field signal after filtering out the pure noise. While for the H$\alpha$ data, each pixel has a line profile instead of a single value. We found that the noise mainly affects the intensity of contrast profiles, but the signal that we are interested in is the Doppler shift shown in the contrast profile. Therefore, our tool is designed to apply the dual-threshold method to the intensity of contrast profiles, and extract Doppler velocities rather than the intensity of interested regions. As a result, our tool constructs a series of marked 3D Doppler maps. These maps are ready for the subsequent feature tracking, which is similar to SWAMIS.

In summary, our RBEs/RREs tracking method is carried out as follows:

i)\ \ \ \ Pre-process the data, including calibration, alignment, bad frame correction, and spatial smoothing;

ii)\ \ \ Construct a normalized reference profile for each pixel, and obtain contrast profiles;

iii)\ \ Apply the dual-threshold feature discrimination algorithm on contrast profiles, and create Doppler velocity maps that are marked with confidence level information;

iv)\ \ \ Employ the general feature tracking procedure, including feature identification and association etc.

It is worth mentioning that different from magnetic features that have either positive or negative polarity, the RBEs and RREs may be overlapped at the same time and location. Thus our method tracks the blue wing for RBEs and the red wing for RREs separately. Our tracking tool records the properties of each detected RBE/RRE, such as central location, shape, horizontal length, and Doppler velocity in each frame during their lifetime.

\section{Results and Discussion}
\label{sect:Result}

\subsection{Observations and Data Processing}
\label{subsect:data}

We check the performance of our tracking tools using the coordinated observations of photospheric LOS magnetograms and chromospheric H$\alpha$ spectral images, as introduced in Section~\ref{sect:Obs}. Here we study a quiet-Sun region near the disk center in the 1 hr period (14:30--15:30 UT on 2011 October 21), when the seeing is relatively better and more stable than the other time periods during our observing run. In order to fully cover the evolution of magnetic features, we analyze the \textit{Hinode}/NFI magnetograms in a wider 2 hr time window (14:00--16:00 UT on 2011 October 21).

We execute a series of preprocessings for both data sets and define proper thresholds. For magnetograms, we follow the steps as outlined in Section~\ref{subsect:smt}. Since NFI data were recorded by two CCD cameras, occasionally there is an offset between them, which is not removed by the standard calibration routine (\verb+fg_prep.pro+ in the solar software). To correct the offset, we use a linear compensation method \citep{Lamb10}. However, there are still some bad frames that show a relatively larger shift of FOV than other frames. To cope with this situation, we first remove each bad frame manually and split the rest images into blocks. For each block, we then average all frames of that block and use it as a reference for alignment. Finally, we align all blocks to the middle block so that the whole image sequence is aligned. In addition, images are Gaussian-smoothed both temporally (using 5 frames) and spatially (using 3~$\times$~3 pixels). After these smoothings, the noise level is about $\sigma=5.5$ G, and we select $2\sigma=11$ G and $3\sigma=16.5$ G as the low and high thresholds of SWAMIS, respectively. We also configure SWAMIS to only track those features that are larger than 4 pixels ($\sim$0$\farcs 1^2$) and appear more than 2 frames ($\sim$2 minutes) in order to minimize the false-positive detection. 

For the IBIS spectral data set, the preprocessing follows a similar procedure as for the magnetogram data. First, we calibrate the data using the standard calibration routines for IBIS. After corrections for the dark and flat fields, each spectral scan is aligned and destretched with the aid of white-light images. The blue shift (especially at the edge of the FOV) induced by the collimated optical setup of IBIS is also compensated. Using a simulated pre-filter curve, we normalize the whole spectra based on the 1 hr data set. As mentioned in Section~\ref{subsect:rbet}, each scan is spatially smoothed using a 3~$\times$~3 Gaussian kernel ($\sigma=1$) and the temporal smoothing is only used to replace bad frames. By checking the fluctuation in contrast profiles, we set the low (high) threshold to 9\% (14\%) absorption level. Finally, we convert the IBIS data set into continuous Doppler maps, and track RBEs and RREs separately using blue-wing and red-wing maps, respectively. Similar to the magnetic cancellation tracking, features smaller than 16 pixels ($\sim$0$\farcs$16$^{2}$) or have a lifetime shorter than 2 frames ($\sim$10~s) are not included for study.

\subsection{Statistical Tracking Result of Magnetic Cancellations}
\label{subsect:re-mc}

\begin{figure}
   \centering
   \includegraphics[width=14.0cm, angle=0]{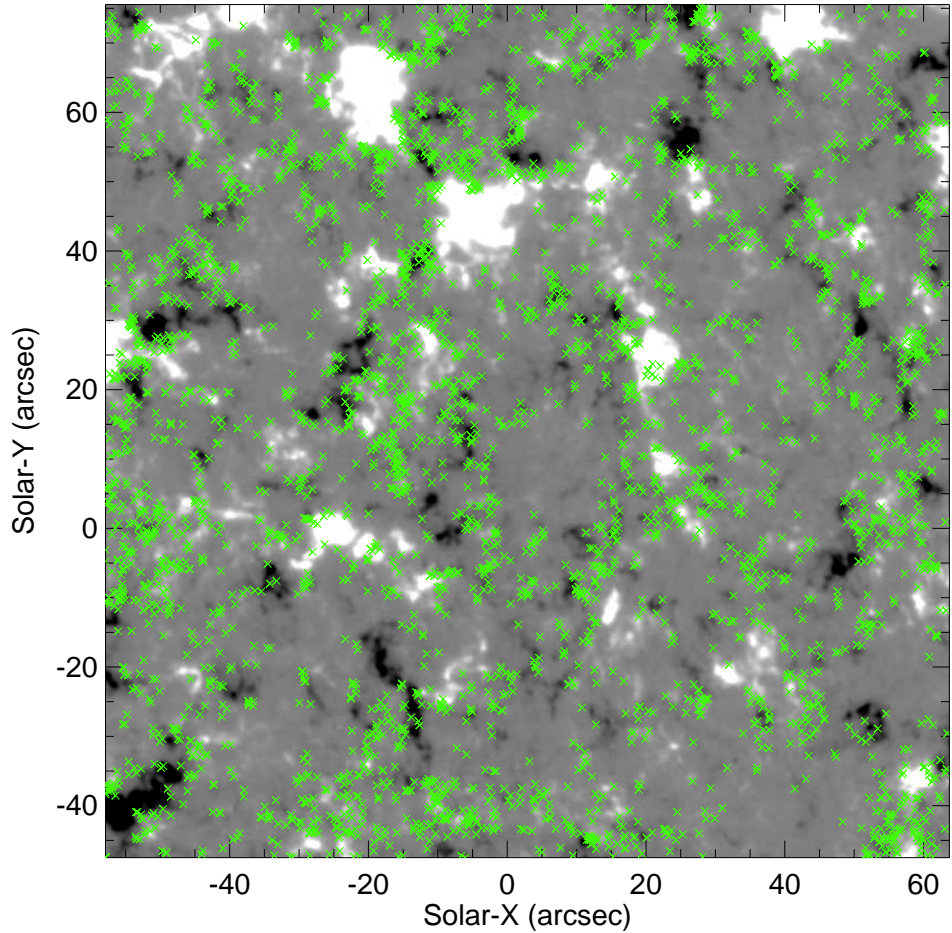}
   \caption{Detected sites of 2969 magnetic cancellation. The background is the average magnetogram in our 2 hr data set (scaled at $\pm$50 G). Our tool tracks the adjacent segments from the border of each opposite-polarity feature and record the center of each pair of adjacent segments as the location of magnetic cancellation (shown as green crosses). Rather than using the middle point between the centers of canceling features, our method provides a more accurate location of cancellation involving a relatively large feature (center of a relatively large feature could be far away from the location of cancellation).} 
   \label{Fig3}
   \end{figure}

\begin{figure}
   \centering
   \includegraphics[width=7.0cm, angle=0]{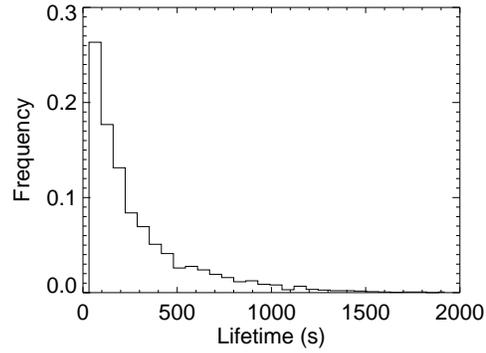}
   \caption{Histogram of lifetime of our detected magnetic cancellations. The cadence of observation is about 64 s.} 
   \label{Fig4}
   \end{figure}
   
\begin{figure}
   \centering
   \includegraphics[width=7.0cm, angle=0]{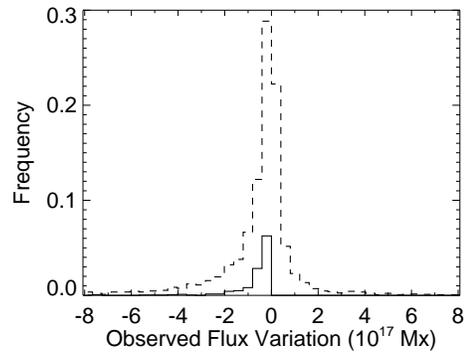}
   \caption{Histogram of the net flux (sum of the signed flux of both canceling features) of our detected 2969 magnetic cancellations (dashed line) covering from birth frame to death frame of each event. The solid line shows the distribution of the most credible 331 cancellations (11.1\%). We note that the net flux variation depends on the feature size and is thus uncertain when tracking magnetic features close to the noise level.} 
   \label{Fig5}
   \end{figure}
   
\begin{figure}
   \centering
   \includegraphics[width=7.0cm, angle=0]{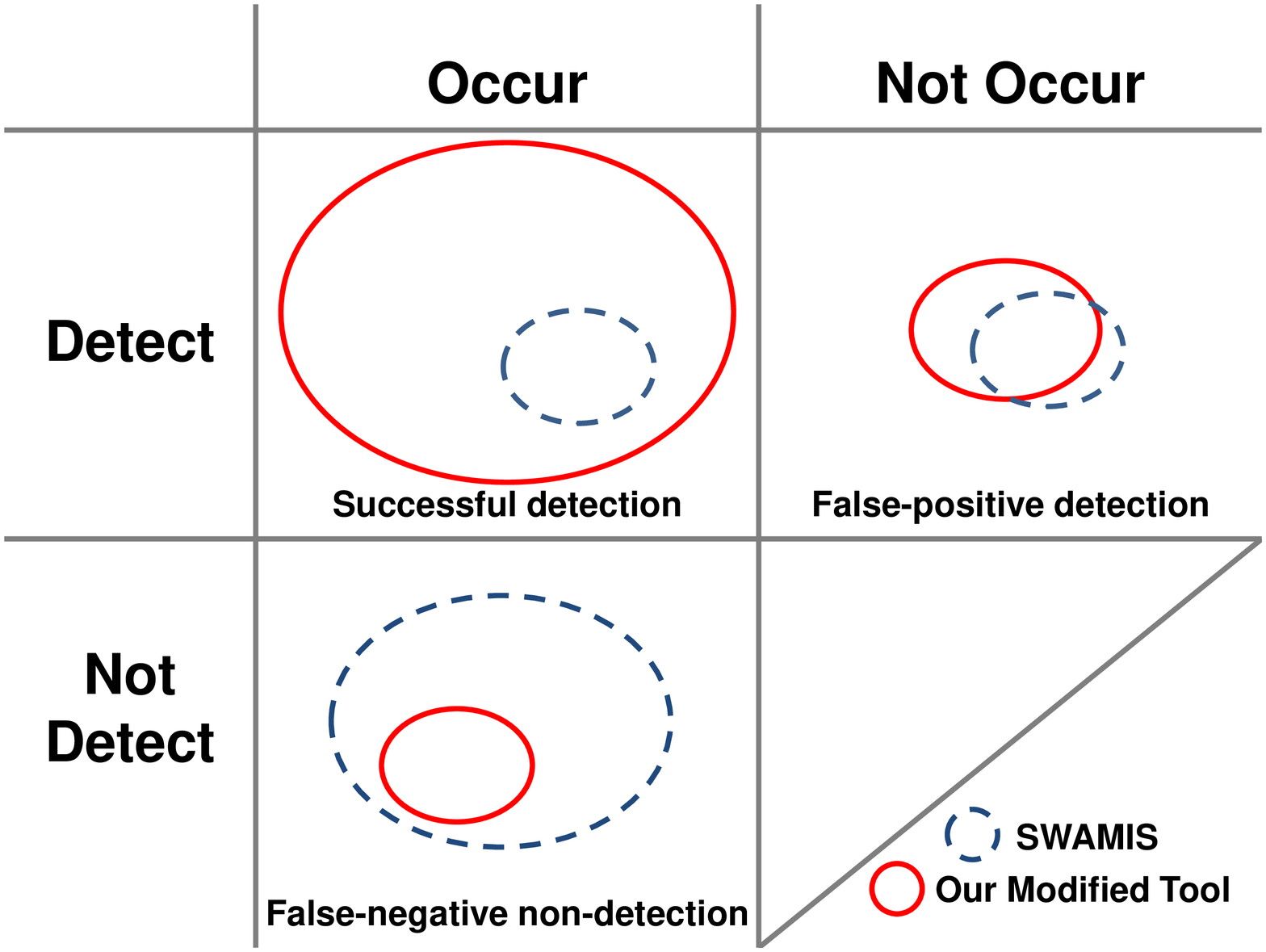}
   \caption{Contingency table of the performance of SWAMIS and our modified tool. Our modified tool detected much more (i.e., eight times) magnetic cancellation with a slightly lower (i.e., around 80\%) credibility comparing to SWAMIS (using data set and thresholds mentioned in Section~\ref{subsect:data}). In other words, our modified tool can significantly reduce false-negative non-detections and keep the false-positive detections in a similar level. Therefore it provides a more comprehensive detection.} 
   \label{Fig6}
   \end{figure}
   
Using our cancellation tracking tool, a total of 2969 magnetic cancellation events are detected during the 2 hr NFI data. This result is further refined based on the credibility that considers the impact of the size and flux balance of features (i.e., credibility greater than 0.75, see Section~\ref{subsect:smt}). We realize that when tracking magnetic features close to the noise level, the flux of a feature could be quite uncertain as errors can accumulate from multiple pixels. Taking this into consideration, only 331 (11.1\%) cancellations appear to reduce the unsigned flux of both positive and negative features during their entire lifetime. Nevertheless, our tool has the ability to identify a large number of possible cancellations and evaluate their credibility based on the surrounding magnetic configuration of features.

Accurate feature locating is another advantage of our cancellation tracking tool. Figure~\ref{Fig3} shows the detected sites of cancellations (green crosses) superimposed on an average magnetogram. It is clear that the cancellations appear to outline the supergranular network as expected. This also implies that the credibility-level evaluation works as planned, in that highly credible cancellations appear at reasonable regions. Therefore, our tool can track more cancellations and pinpoint their accurate locations, and at the same time still minimizes the false-positive detection.

Based on the particular condition of our observation and the used thresholds, we statistically study the distribution of magnetic cancellations. The extrapolated occurrence rate of magnetic cancellation over the whole Sun is 80$\pm$33~s$^{-1}$ ($1-\sigma$; 64 s cadence). As shown in Figure~\ref{Fig4}, the lifetime shows a monotonic decrease with a mean of 5 minutes. There are several studies analyze the flux cancellation rate in the photosphere \citep{Chae02,Park09}, however it may not be appropriate for our study due to the following reason. In Figure~\ref{Fig5}, we show the distribution of the total net flux variation of both canceling features. SWAMIS relies more on this property; however we find that there are some cancellations with high credibility that exhibit an increase of the total net flux. Although a magnetic cancellation is generally defined as a decrease of the total net flux, it can be easily misidentified due to the error of observation and detection. As mentioned in Section~\ref{subsect:smt}, we found that the size difference of canceling features plays an important role thus can not be ignored when checking the net flux balance. Figure~\ref{Fig6} illustrates the different performances of SWAMIS and our modified tool.

\subsection{Statistical Tracking Result of RBEs}
\label{subsect:re-rbe}

\begin{figure}
   \centering
   \includegraphics[width=7.0cm, angle=0]{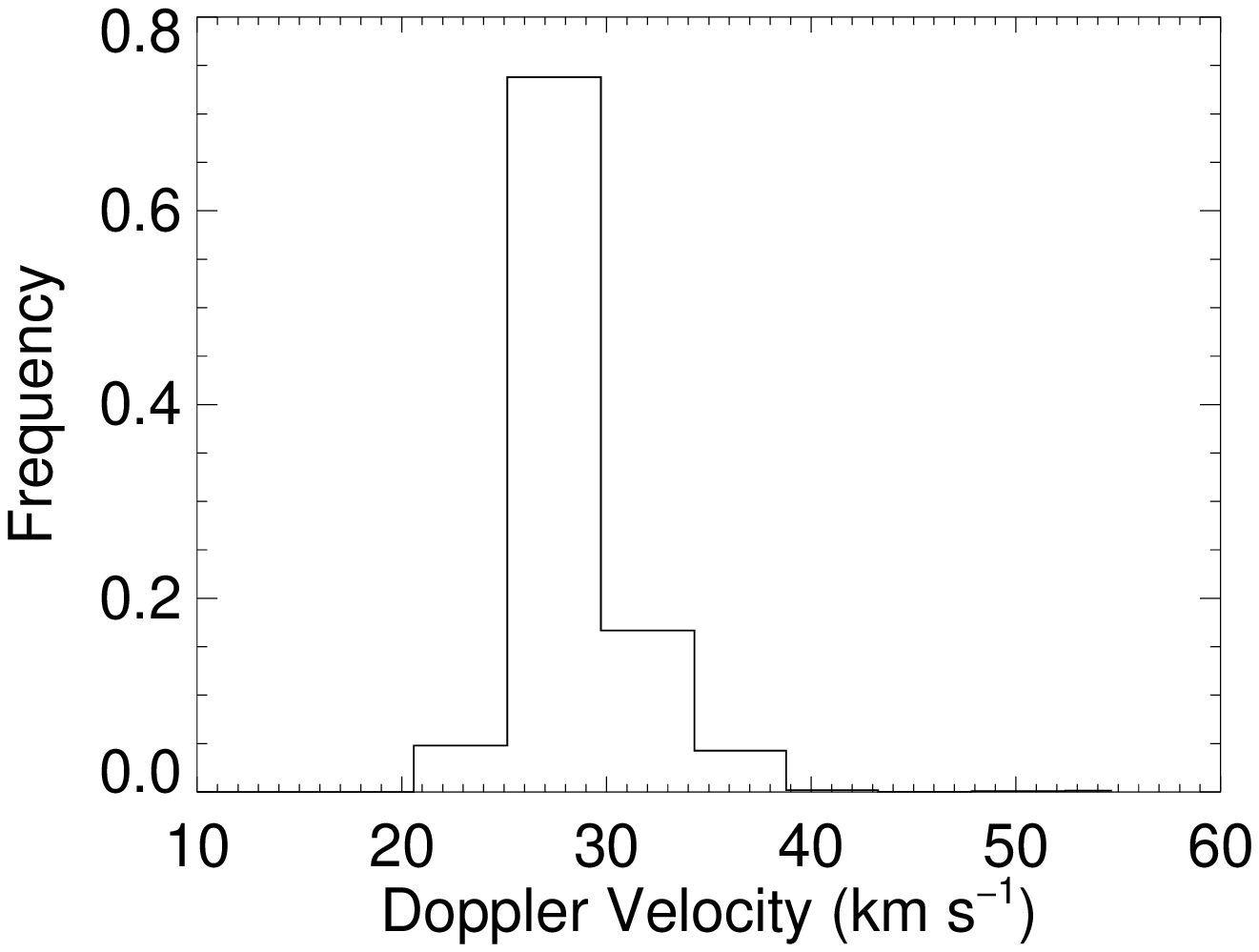}
   \caption{Histogram of Doppler velocity of our detected RBEs. For each RBE, we use the peak velocity during its lifetime.} 
   \label{Fig7}
   \end{figure}
   
\begin{figure}
   \centering
   \includegraphics[width=7.0cm, angle=0]{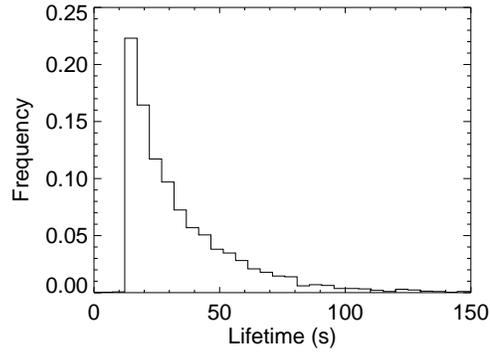}
   \caption{Histogram of the life time of our detected RBEs. The data cadence is $\sim$4.85 s. The cut-off at $\sim$14.5 s (3 frames) is an artifact caused by our tracking method.} 
   \label{Fig8}
   \end{figure}

\begin{figure}
   \centering
   \includegraphics[width=7.0cm, angle=0]{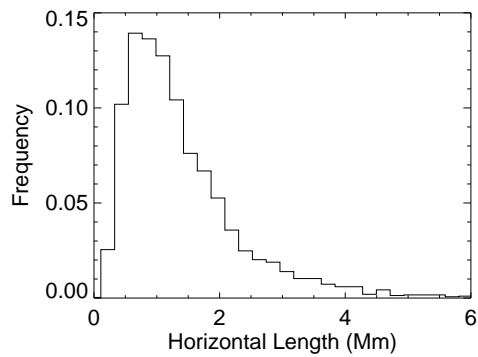}
   \caption{Histogram of the horizontal length of RBEs. We use the maximum length for each RBE.} 
   \label{Fig9}
   \end{figure}
   
Using our 1 hr IBIS data set, we found 3022 RBEs that are not paired with RREs. Considering that an RBE is defined as a feature that shows no red-shifted component, we removed features detected from the blue wing, which are temporally and spatially associated or close to any feature detected from the red wing, as they could be the disk counterparts of type I spicules (with both upflow and downflow). Although \citet{Sekse13+} interpreted this situation using a combination of transverse and torsional motions of RBEs under specific viewing angles, it is still under debate \citep{Lipartito14}. In our data set, we hardly found any paired RBEs/RREs that are parallel to each other indicating a torsional tube of plasma.

Figure~\ref{Fig7} shows the histogram of Doppler velocity of the 3022 RBEs. The Doppler velocity mainly ranges from 20 to 40 km~s$^{-1}$ ($\sigma=3.0$ km~s$^{-1}$), with a mean value of 28.5 km~s$^{-1}$. This result is comparable to previous studies \citep{Wang98,RvdV09,Sekse13} but our mean velocity is slightly higher, which could be related to the different thresholds used. We note that there does not exist a clear cut-off Doppler velocity of RBEs, as also mentioned in \citet{Sekse13}, since decreasing the detection threshold would lead to more RBEs. Taking advantage of the dual-threshold method that utilizes the whole spectral profile, our tool can reveal RBEs that has a low Doppler velocity but still exhibit a credible blue-shifted component. Furthermore, we find the lifetime of RBEs (Figure~\ref{Fig8}) is 35.2$\pm$27.5 s ($1-\sigma$; data cadence is $\sim$5 s), as well as derive the occurrence rate of RBEs extrapolated to the whole Sun, which turn out to be 331$\pm$292 s$^{-1}$. These results agree with previous studies \citep{RvdV09,Sekse13}. In addition, the horizontal length of the detected RBEs is found to be 1.50$\pm$0.96 Mm (Figure~\ref{Fig9}), and the shape of RBEs can be elongated (49\%), round (22\%), or irregular (29\%). The RBEs in our observation are shorter in length and less elongated compared to previous studies mentioned above. This is probably due to the lack of the projection effect as our target region is much closer to the disk center.
 
\subsection{Relation Between RBEs and Magnetic Cancellations}
\label{subsect:relation}

\begin{figure}
   \centering
   \includegraphics[width=13.7cm, angle=0]{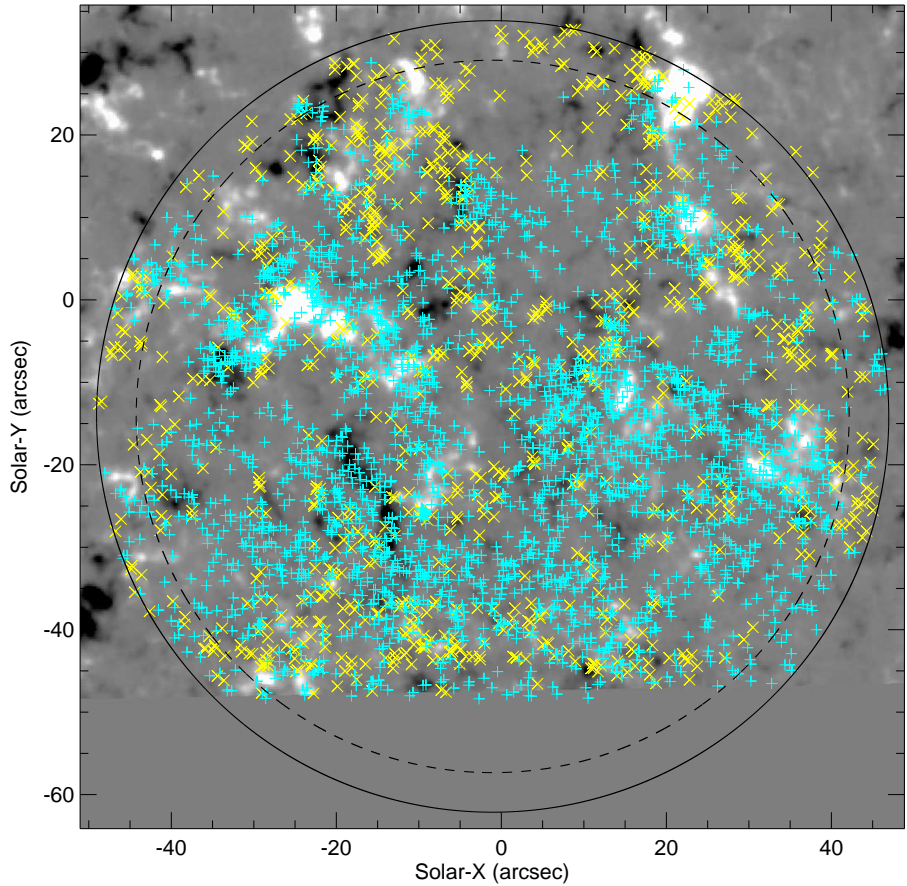}
   \caption{Spatial distributions of our detected 936 magnetic cancellations (yellow crosses) and 2715 RBEs (cyan crosses), superimposed on an average magnetogram from 14:00 to 16:00 UT (scaled at $\pm$50 G). The NFI magnetogram has been aligned with IBIS. The FOV of IBIS is denoted by the solid circle, while the dashed circle shows the central region with a 0.9 radius of the IBIS FOV. At the top edge of the IBIS FOV, RBEs are not identified due to a calibration problem.} 
   \label{Fig10}
   \end{figure}
   
Combining the tracking results of magnetic cancellations and RBEs, we are attempt to look at the possible correlation between them using our coordinated observations. After a careful image alignment using the intensity images of NFI and white-light images of IBIS, we present in Figure~\ref{Fig10} the spatial distribution of the both identified features. It can be seen that the detected locations of magnetic cancellations and RBEs show some patterns rather than a uniform random distribution. As mentioned earlier, magnetic cancellations tend to concentrate on the magnetic network boundary, and indeed, most of them are surrounded by RBEs in our result as expected. Inside the network magnetic cancellations are much less frequent. Accordingly, there are much less RBEs as well. In addition, there are several unipolar regions where many RBEs are present but are lack of magnetic cancellations. 

As for the temporal association, it is extremely difficult to establish possible correlation in timing between RBEs and magnetic flux cancellation. Nevertheless, this does not imply that these two phenomena are totally independent of each other. In fact, we found that it could be technically difficult to examine the temporal correlation between them. Since RBEs occur frequently with a lifetime shorter than a minute and some magnetic cancellations could last 30 times longer than RBEs, it is always be able to associate a nearby magnetic cancellation for a given RBE. Therefore, it is very hard to objectively study the correlation between these two kinds of evolving features.

Finally, the statistical relation between magnetic cancellations and RBEs seems to be vague, as (1) both activities show a wide variety of their properties. (2) Both detection methods of cancellations and RBEs are not completely accurate, considering that either false-positive detection or false-negative non-detection may further increase the variance of their properties. (3) The confidence interval of correlation is more than a linear superposition of the variance of event properties. As a result, the confidence interval of correlation could easily become broad, which makes it less reliable. In this case, for example, even if every magnetic cancellation is related to an RBE, their correlation (either spatial or temporal) would not be close to 100\% based on the detected features. To solve these problems, clearer definitions of magnetic cancellations and RBEs would be helpful, as they could reduce the variance of feature properties by limiting the samples themselves as well as increasing the accuracy of detection.

\section{Conclusions}
\label{sect:Con}

In summary, we have developed two automatic feature tracking tools for magnetic flux cancellations and RBEs. For the cancellation tracking, we develop a dedicated algorithm, which uses the intermediate results of SWAMIS as input and can provide more accurate results compared to the standard SWAMIS code on the same magnetograms. For RBEs tracking, our tool is able to detect RBEs in H$\alpha$ spectral images even if slight image distortion and nonuniform background are present.

Our tools are functioning well when applied to our coordinated observations of \textit{Hinode}/NFI and NSO/IBIS. Magnetic cancellations detected by our tool are concentrated on the magnetic network boundaries. Statistically, our results show similar properties of RBEs compared to previous studies \citep{Wang98,RvdV09,Sekse13}. Furthermore, we investigate the potential relation between magnetic cancellations and RBEs. We find that magnetic cancellations and RBEs are spatially correlated to the certain extent; however, their temporal correlation is hard to be established, due to very frequent occurrence of RBEs and long lasting magnetic flux cancellation.

\begin{acknowledgements}
Observations presented in this study were partially obtained with the facilities of NSO, which is operated by the Association of Universities for Research in Astronomy, Inc. (AURA), under cooperative agreement with the National Science Foundation. IBIS has been built by the INAF/Osservatorio Astrofisico di Arcetri with contributions from the Universities of Firenze and Roma Tor Vergata, the NSO, and the Italian Ministries of Research (MUR) and Foreign Affairs (MAE). This research has made use of the IBIS software reduction package provided and made available by the NSO (special thanks goes to Dr. Alexandra Tritschler). The authors thank the NSO observers (Doug Gilliam, Joe Elrod, and Mike Bradford) for their professional and excellent observation support. The authors also thank the referee for valuable comments which help us to improve this paper. \textit{Hinode} is a Japanese mission developed and launched by ISAS/JAXA, with NAOJ as domestic partner and NASA and UKSA as international partners. It is operated by these agencies in co-operation with ESA and NSC (Norway). This research was supported by NASA under grants NNX11AO70G, NNX13AF76G and NNX14AC12G and by NSF under grant AGS 1153226, 1250374, 1348513 and 1408703. D.A.L. was partially supported by NASA grants NNX08AJ06G and NNX11AP03G. S. Park was supported through the project "SOLAR-4068" which is implemented under the "ARISTEIA II" Action of the operational programme "Education and Lifelong Learning" and is co-funded by the European Social Fund (ESF) and National funds.
\end{acknowledgements}

\bibliographystyle{raa}
\bibliography{xin2014raa}


\end{document}